# High Critical Current Density 4 MA/cm$^2$ in Co-Doped BaFe$_2$As$_2$ Epitaxial Films Grown on (La, Sr)(Al, Ta)O$_3$ Substrates without Buffer Layers

Takayoshi Katase[1*], Hidenori Hiramatsu[2], Toshio Kamiya[1], and Hideo Hosono[1,2]

[1]Materials and Structures Laboratory, Mailbox R3-1, Tokyo Institute of Technology, 4259 Nagatsuta, Midori, Yokohama 226-8503, Japan

[2]Frontier Research Center, S2-6F East, Mailbox S2-13, Tokyo Institute of Technology, 4259 Nagatsuta, Midori, Yokohama 226-8503, Japan

**ABSTRACT**

High critical current densities $J_c$ of 4 MA/cm$^2$ at 4 K were obtained in Co-doped BaFe$_2$As$_2$ (BaFe$_2$As$_2$:Co) epitaxial films grown directly on (La, Sr)(Al, Ta)O$_3$ substrates by pulsed laser deposition. Use of a highly pure target and improvement of film homogeneity were the critical factors to achieve the high $J_c$. The improved BaFe$_2$As$_2$:Co epitaxial films contained almost no Fe impurity and showed high crystallinity (crystallite tilt angle $\Delta\omega = 0.5$ ° and twist angle $\Delta\phi = 0.5$ °) and a sharp superconducting transition with a width of 1.1 K. It is considered that these improvements resulted in the enhanced superconducting properties comparable to those of single crystals.

[*]E-mail address: katase@lucid.msl.titech.ac.jp

Iron-based superconductors[1] possess attractive properties such as high critical temperatures ($T_c$) up to 56 K[2] and high upper critical magnetic fields > 100 T[3], which arouses active research to grow high-quality epitaxial films[4] in order to examine intrinsic properties and grain boundary properties, and to clarify the potential for device applications. In such circumstances, epitaxial film growth of Fe-based superconductors such as fluorine-doped $Ln$FeAsO ($Ln$ = La, Nd)[5-8], cobalt-doped $AE$Fe$_2$As$_2$ ($AE$ = Sr, Ba)[9-15] and Fe(S, Se, Te)[16-29] have been reported to date. Among them, cobalt-doped BaFe$_2$As$_2$ (BaFe$_2$As$_2$:Co) epitaxial films have received much attention because of relatively easy growth by pulsed laser deposition (PLD) and high stability even in a moist atmosphere[10], which are distinct advantages over $Ln$FeAsO[4-8,30].

However, in our first report, the BaFe$_2$As$_2$:Co epitaxial films grown on (La, Sr)(Al, Ta)O$_3$ (LSAT) substrates showed low critical current densities ($J_c$) of ~$10^5$ A/cm$^2$, although these values were larger than those reported for cobalt-doped SrFe$_2$As$_2$ epitaxial films[31]. For an application to Josephson junctions and an exploration of grain boundary properties, it is necessary to improve crystal quality and superconducting properties. In particular, $J_c$ over $10^6$ A/cm$^2$ is an important requirement for fabricating Josephson junctions.

Lee *et al.*[11] and Iida *et al.*[13] reported that higher $T_c$ and $J_c$ were obtained for BaFe$_2$As$_2$:Co films grown on SrTiO$_3$ (STO) substrates; they claimed that these films had better quality than those on LSAT substrates. However, a STO substrate becomes highly conductive if it is used for film deposition under a reductive condition including a high vacuum, which makes carrier transport measurements difficult. It is, therefore, necessary to grow high-quality BaFe$_2$As$_2$:Co films on insulating substrates, although Lee *et al.* concluded that the use of buffer layers such as (001) STO and BaTiO$_3$ was necessary to obtain a high $J_c$ of ~5 MA/cm$^2$ at 4.2 K[14] if a LSAT substrate was used.

In this study, we report that high $J_c$ of 4 MA/cm$^2$ at 4 K are obtained in BaFe$_2$As$_2$:Co epitaxial films even if grown directly on LSAT substrates without buffer layers. The key factor to fabricate the high-quality BaFe$_2$As$_2$:Co epitaxial films is extreme optimization of a film fabrication process; e.g., improving the phase purity of the BaFe$_2$As$_2$:Co PLD target and the homogeneity of the substrate temperature. Very recently, we reported fabrication of Josephson junctions using the high-$J_c$ BaFe$_2$As$_2$:Co epitaxial films[32].

We reported growth of BaFe$_2$As$_2$:Co epitaxial films[10], which is the first fabrication of superconducting BaFe$_2$As$_2$:Co epitaxial films; however, these films contained several % of Fe impurities and showed low onset $T_c$ ($T_c^{onset}$) ~ 20 K, broad superconducting transition widths ($\Delta T_c$) ~ 3 K, and low $J_c$ ~ 0.1 MA/cm$^2$, which were not satisfactory to develop superconducting devices such as Josephson junctions. We, therefore, carefully tuned the film fabrication process to reduce the Fe impurity and improve the film uniformity. Furthermore, we decreased the Co content in BaFe$_{2-x}$Co$_x$As$_2$ PLD targets from $x$ = 0.2 to 0.16, the latter of which is the optimal content for BaFe$_{2-x}$Co$_x$As$_2$ single crystals with $T_c$ of 22 K[33].

PLD targets were synthesized by solid-state reactions of a stoichiometric mixture of BaAs, Fe$_2$As, and Co$_2$As via a reaction BaAs + 0.92Fe$_2$As + 0.08Co$_2$As $\rightarrow$ BaFe$_{1.84}$Co$_{0.16}$As$_2$. All the processes except for heating were performed in an argon-filled glove box. BaAs was synthesized by a reaction of Ba and As (Ba:As = 1:1), which was conducted in an evacuated silica-tube by heating a mixture of Ba metal (99.99 wt%) and gray As powder (99.9999 wt%) at 700 °C for 10 h. Single-phase Fe$_2$As and Co$_2$As were obtained by heating mixtures of (Fe, As) and (Co, As), respectively, at 850 and 900 °C for 10 h (the purity of Fe powder was 99.9 wt% and that of Co powder 99 wt%). These intermediate ingredients were then mixed at the stoichiometric ratio of BaFe$_{1.84}$Co$_{0.16}$As$_2$, pressed, and heated in an evacuated silica-tube at 900 °C for 16 h to obtain a sintered pellet. Phase purity of the resulting PLD target was examined by X-ray diffraction (XRD, anode radiation: CuK$\alpha$, D8 ADVANCE-TXS, Bruker AXS).

BaFe$_2$As$_2$:Co epitaxial films were deposited on mixed perovskite (La, Sr)(Al, Ta)O$_3$ (LSAT) (001) single-crystal substrates by ablating the BaFe$_2$As$_2$:Co PLD target using a second harmonic (wavelength: 532 nm) of a pulsed Nd:YAG laser in a vacuum at ~10$^{-5}$ Pa. In order to improve the homogeneity of the substrate temperature, the LSAT substrates were tightly fixed to a substrate carrier (inconel alloy) so as to make a good thermal contact.

Film structures including crystalline quality and crystallites orientation were examined by high-resolution XRD (HR-XRD, anode radiation: CuK$\alpha_1$, ATX-G, Rigaku) at room temperature. Temperature ($T$) dependences of electrical resistivity ($\rho$) were measured by the four-probe method in the temperature range of 2 – 305 K with a physical property measurement system (PPMS, Quantum Design). $J_c$ were estimated from current ($I$) – voltage ($V$) curves using 300-µm-long and 10-µm-wide micro-bridges with a 1 µV criterion.

In the previous paper[10], the BaFe$_2$As$_2$:Co PLD target contained several % of impurities such as Fe, FeAs, Fe$_2$As, and FeAs$_2$, indicating that the chemical composition of the target deviated from the stoichiometry with less amounts of Ba and As. Therefore, we optimized the preparation process of BaAs. Figure 1 shows the optimized procedure for the BaAs synthesis. First, large pieces of Ba metal were finely cut [(a) → (b)], and then crushed into small planar grains [(b) → (c)]. This procedure was repeated until the piece sizes became ~ 1 mm [(c) → (d)], which is crucial to enhance the reaction between Ba metal with As powder. The obtained BaAs powders looked black and did not contain residual Ba metal (e).

Figure 2 shows XRD patterns of the improved BaFe$_2$As$_2$:Co PLD target obtained by the optimized process (top panel) and the previous PLD target (bottom panel). The inset figure shows magnified XRD patterns in the range of $2\theta = 50^\circ - 70^\circ$. The previous PLD target contained impurities of Fe (indicated by the purple line), Fe$_2$As (the green line), FeAs$_2$ (the blue lines), and FeAs (the red lines). On the other hand, the Fe impurity was not detected and the iron arsenide impurities were almost removed in the improved PLD target. These results indicate that it is important to use finely-cut Ba metal to enhance the formation rate of the BaAs, which leads to a single-phase BaFe$_2$As$_2$:Co PLD target.

Figure 3(a) shows out-of-plane XRD patterns of the the previously reported BaFe$_2$As$_2$:Co epitaxial films (low-quality LQ, bottom panels) and the improved BaFe$_2$As$_2$:Co epitaxial films deposited using the high-purity PLD target (high-quality HQ, top panels). The LQ films exhibited intense peaks of BaFe$_2$As$_2$:Co 00$l$ and LSAT 00$l$ diffractions along with an extra peak at $2\theta = 65^\circ$, which can be assigned to the 200 diffraction of Fe metal. On the other hand, the HQ films showed only the BaFe$_2$As$_2$:Co 00$l$ and LSAT 00$l$ diffractions. Figures 3(b) and 3(c) show (b) out-of-plane rocking curves of the 002 diffractions and (c) in-plane rocking curves of the 200 diffractions of the HQ and LQ BaFe$_2$As$_2$:Co films, respectively. The full width at half maximum (FWHM) Δω of the 002 rocking curve was 0.5 degrees for the HQ film, which was slightly sharper than that of the LQ film (0.6 degrees). In addition, the FWHM Δ$\phi$ of the 200 rocking curve of the HQ film was 0.5 degrees, which was approximately one third of that of the LQ film (1.8 degrees). These results indicate that crystal quality and crystallites orientation were improved significantly for the HQ films. By contrast, the LQ films exhibited visually-observable inhomogeneity, which would be the major reason for the broadened rocking curves.

Figure 4(a) shows $\rho$–$T$ curves for the BaFe$_2$As$_2$:Co epitaxial films. The $\rho$–$T$ curve of the HQ film is indicated by the red circles, and that of the LQ film by the blue squares. The LQ film showed a low $T_c^{onset}$ of 20.2 K and a wide $\Delta T_c$ of 2.9 K. On the other hand, the HQ film showed a higher $T_c^{onset}$ of 22.6 K and a sharper $\Delta T_c$ = 1.1 K, showing that the superconducting properties of the HQ film are similar to those of single crystals[34]. The sharp $\Delta T_c$ of the HQ film reflects high homogeneity of the superconducting properties.

Figure 4(b) shows temperature dependences of critical current density of the HQ film and the LQ film measured using the 10μm-wide micro-bridges. The $J_c$ at $T$ = 4 K was 4.0 ×10$^6$ A/cm$^2$ for the HQ film and 2.1 ×10$^5$ A/cm$^2$ for the LQ film. This $J_c$ value is much higher than those of recently-reported high-quality BaFe$_2$As$_2$:Co epitaxial films grown on LSAT substrates[11,15] and comparable to that of the high $J_c$ films using STO buffer layers. It is evident from these results that the improvement of phase purity and crystal quality of BaFe$_2$As$_2$:Co epitaxial films widens a process window for growing higher-quality epitaxial films and allows us to use a wider variety of substrates, leading to the enhanced superconducting properties. Note that recent reports suggest another origin for high $J_c$ in BaFe$_2$As$_2$:Co epitaxial films [14,35]. Lee and co-workers found vertically-aligned line defects composed mainly of a Ba-Fe-O secondary phase in their BaFe$_2$As$_2$:Co epitaxial films grown on STO buffer layers and suggested that these defects resulted in strong $c$-axis pinning of vortex, which would explain why $J_c$ of their and our BaFe$_2$As$_2$:Co films are much larger than that of single crystals (self-field $J_c$ of 0.4 MA/cm$^2$ at 4.2 K)[36]. As it is difficult to completely deplete oxygen-related impurities from a vacuum deposition chamber, this possibility remains also for our case. Detailed microstructure analyses are necessary to verify the origin of the high $J_c$.

In summary, we demonstrated that high $J_c$ of 4 MA/cm$^2$ at 4 K were obtained in BaFe$_2$As$_2$:Co epitaxial films grown directly on LSAT substrates. The high-quality BaFe$_2$As$_2$:Co films were obtained by using the highly-pure BaFe$_2$As$_2$:Co PLD target and homogenizing the substrate temperature, which enhanced the superconducting properties including $J_c$. We actually succeeded in fabricating Josephson junctions using the high-quality BaFe$_2$As$_2$:Co films as reported in ref. 32. The present achievement will also develop other superconducting devices such as superconducting quantum interference devices (SQUIDs). It would also encourage research of heteroepitaxial growth on various substrates including MgO and LaSrAlO$_4$ with low dielectric constants, which would be a key to fabricate high-frequency superconducting devices.

**Acknowledgment**

This work was supported by the Japan Society for the Promotion of Science (JSPS), Japan, through "Funding Program for World-Leading Innovative R&D on Science and Technology (FIRST) Program".

**Figure captions**

Fig. 1. Optimized preparation procedure of BaAs starting material. (a) Ba metal, distilled, dendritic pieces (Aldrich). (b) Finely-cut Ba metal pieces. (c) Crushed plates of Ba metal. (d) Fine Ba plates ~1 mm in size. (e) Completely reacted BaAs in an evacuated silica-tube.

Fig. 2. XRD patterns of the previously obtained $BaFe_{1.8}Co_{0.2}As_2$ PLD target (top panel) and the improved $BaFe_{1.84}Co_{0.16}As_2$ PLD target (bottom panel). The inset figures show magnified patterns in the range of $2\theta = 50^\circ - 70^\circ$.

Fig. 3. (a) Out-of-plane XRD patterns of $BaFe_2As_2$:Co films. (b) Out-of-plane rocking curves ($\omega$ scans) of the 002 diffractions and (c) in-plane rocking curves ($\phi$ scans) of the 200 diffractions. Bottom and top panels show those of the previously reported low-quality (LQ) film and the improved high-quality (HQ) films, respectively.

Fig. 4. (a) $\rho - T$ curves and (b) $J_c - T$ curves of the HQ epitaxial films (red circles) and LQ epitaxial films (blue squares), respectively.

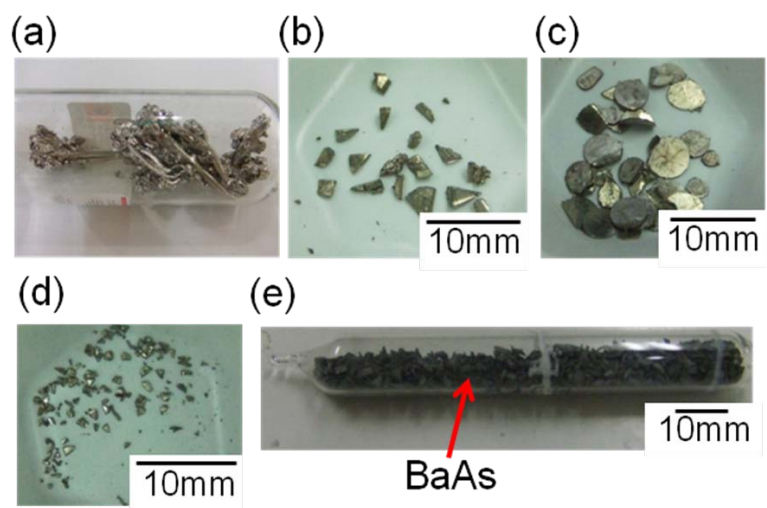

Fig.1

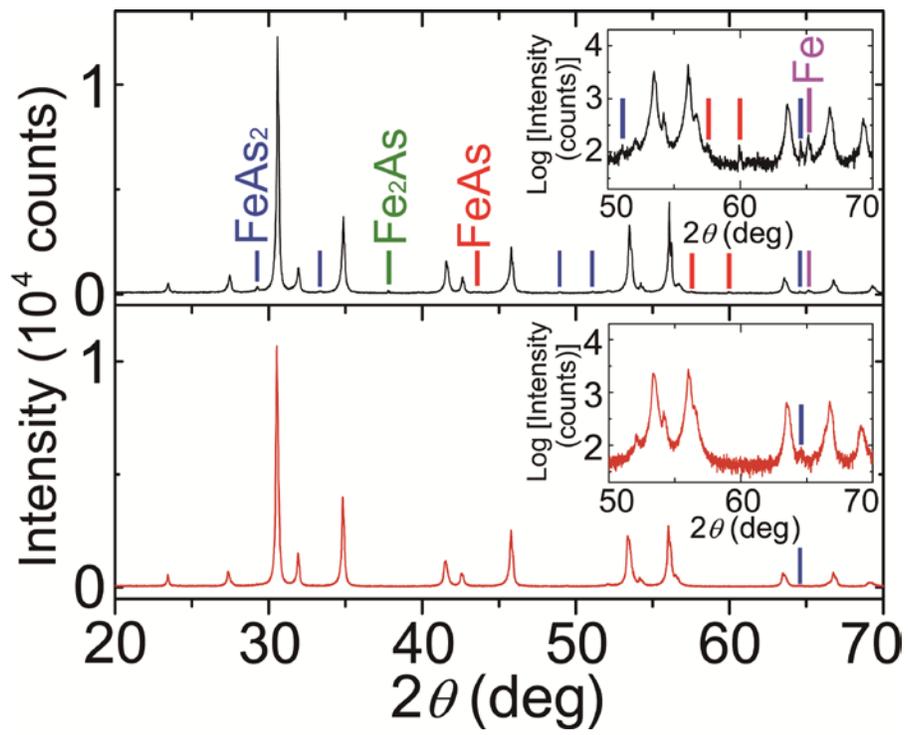

Fig.2

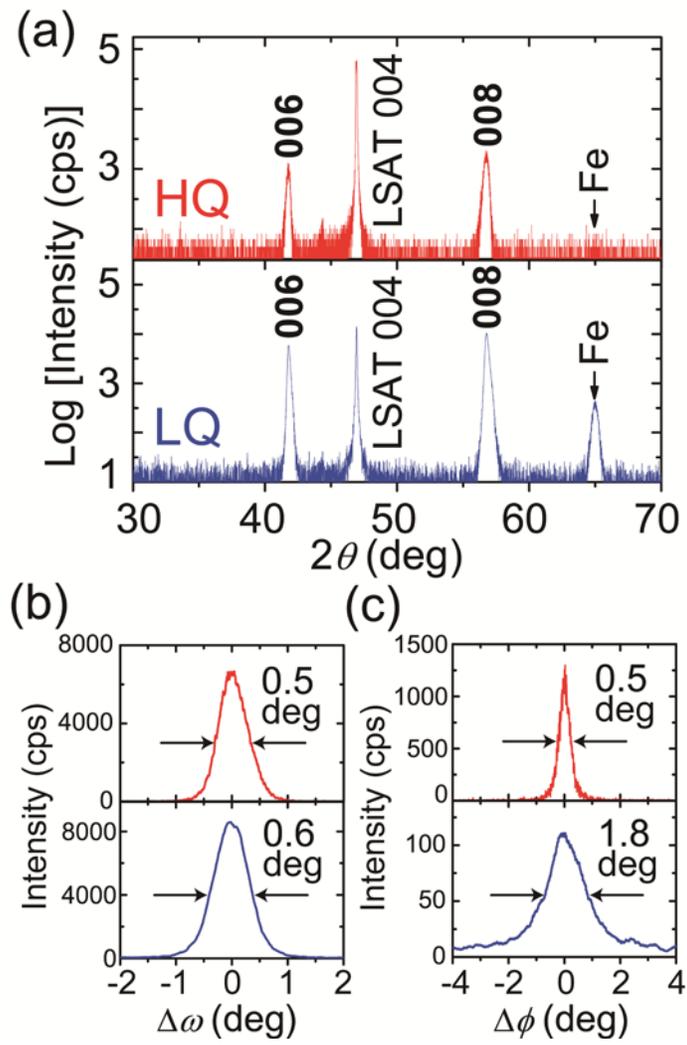

Fig.3

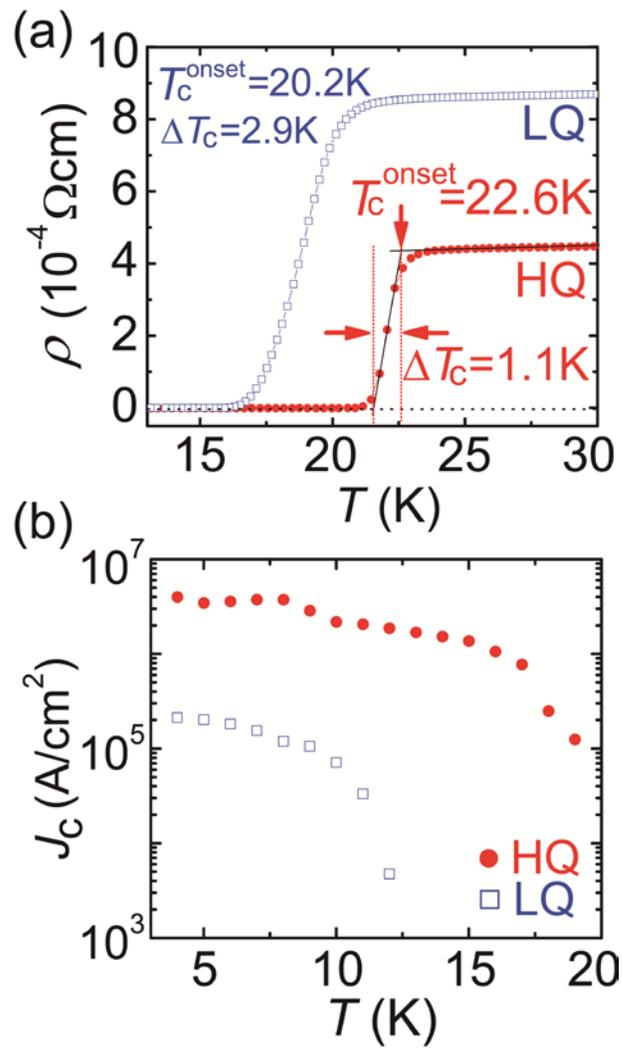

Fig.4